\documentstyle[aps,preprint]{revtex}

\begin{document}
\draft
\title{Controlled entanglement of two field modes in a Cavity Quantum Electrodynamics experiment}
\author{A. Rauschenbeutel, P. Bertet, S. Osnaghi, G. Nogues, M. Brune, J.M. Raimond, S. Haroche}
\address{Laboratoire Kastler Brossel\
\thanks{Unit\'e mixte de recherche, Ecole normale sup\'erieure, Universit\'e Pierre 
et Marie Curie et CNRS (UMR8552)},\\
 D\'epartement de Physique
de l'Ecole Normale Sup\'erieure,\\
24 rue Lhomond, F-75231 Paris Cedex 05, France\\
}
\date{\today}
\maketitle
\begin{abstract}
Using a single circular Rydberg atom, we have prepared two modes of a superconducting 
cavity in a maximally entangled state. The two modes share a single photon. This 
entanglement is revealed by a second atom probing, after a delay, the correlations 
between the two modes. This experiment opens interesting perspectives for quantum information manipulation and fundamental tests of quantum theory.
\end{abstract}
\vskip2pc

The preparation of complex quantum entangled states in well-controlled conditions 
is the subject of an intense experimental activity. The manipulation of these states 
which have non-classical and non-local properties leads to a better understanding of 
basic quantum phenomena. Complex entangled states, such as the Greenberger, Horne 
and Zeilinger triplets of particles \cite{GHZ} are used for tests of quantum non-locality 
\cite{GHZEXP}. Moreover, the relaxation dynamics of large entangled states sheds light on 
the decoherence process and on the quantum-classical boundary \cite{ZUREK}. 

Beyond these fundamental aspects, entangled quantum states can be used for information 
processing. Elements of information may be coded on quantum two-level systems (``qubits") 
\cite{GATES}. Quantum entanglement between qubits leads to new functions for information 
processing or transmission. Non-local correlations between two qubits can be used for 
quantum key distribution \cite{EKERT} or teleportation \cite{BENNETT}. More complex 
entanglement manipulations could be used for quantum error correction \cite{STEANE} 
or entanglement purification \cite{ZOLLER}. The manipulation of controlled entangled 
states, protected from their environment, is experimentally challenging. Clear-cut 
entanglement between individual systems has been achieved so far only in quantum optics, 
in photon down-conversion processes \cite{GHZEXP}, with trapped ions \cite{WINELAND} 
or in cavity quantum electrodynamics \cite{STEP}. 

In the latter case, the entanglement results from the interaction of a two-level atom 
with a cavity field mode. With circular Rydberg atoms and superconducting cavities, 
the coherent atom-field coupling overwhelms dissipation \cite{ADVANCES2}. The basic 
interaction process is the vacuum Rabi oscillation \cite{QRABI}. An atom, entering 
the empty cavity in the upper state of the transition resonant with the mode, 
periodically emits and absorbs a photon. Interrupting the coherent interaction 
at selected times, one can generate atom-cavity entanglement \cite{EPRPAIR}, 
exchange atom and cavity states \cite{MEMORY} or realize a two-qubit quantum 
logic gate \cite{QGATE}, basic unit of quantum information processing systems. 
These functions can be arranged to realize a quantum non-demolition measurement 
of a single photon \cite{QND} or to generate a three-particle entangled state \cite{STEP}.

Up to now, these experiments have involved a single cavity mode interacting with 
one, two or three atoms crossing it successively. We present here an important 
step towards the realization of more complex entanglement in cavity QED, by involving 
in the experiment two independent cavity modes. We have shown that a single atom 
can be used to entangle the two modes, a second atom being employed later to reveal 
this entanglement. This multi-mode entanglement opens new perspectives for quantum 
information processing and non-locality tests.

The principle of the experiment is sketched on figure \ref{FIG_SCHEME}(a). Circular 
Rydberg atoms cross, one at a time, the superconducting cavity $C$ which sustains 
two non-degenerate orthogonally polarized modes $M_a$ and $M_b$, initially in 
the vacuum state $|0\rangle$. The $M_a-M_b$ frequency difference is $\delta$ 
($M_a$ has the highest frequency). An experimental sequence involves two circular 
Rydberg atoms $A_s$ and $A_p$ with two levels $e$ and $g$ (upper and lower states 
respectively). $A_s$ is a source atom which emits a single photon coherently 
shared between the two cavity modes at different frequencies. Atom $A_p$ probes 
at a later time this single photon state. The final states of both atoms ($e$ 
or $g$) are analyzed in the state-selective field-ionization detector $D$. 

The detuning $\Delta$ between the frequency of the  $e\rightarrow g$ transition 
and mode $M_a$ can be tuned through Stark effect. When $\Delta=0\ (-\delta)$, 
the atom is in resonance with mode $M_a$ ($M_b$). When the atom is tuned to 
resonance with one mode, the interaction with the second one has a small effect, 
provided that $\delta$ is much larger than the vacuum Rabi oscillation frequency 
$\Omega$. Finally, the detuning $\Delta$ can also be set to a large negative 
value, freezing the atom-field evolution.

The experiment timing is schematized on figure \ref{FIG_SCHEME}(b). Atom $A_s$ 
enters $C$ in level $e$ at time $t=0$. The initial $A_s-M_a-M_b$ state is 
$|e_s,0_a,0_b\rangle$. Atom $A_s$ interacts first resonantly with  $M_a$ for 
a $\pi/2$ spontaneous emission pulse \cite{QRABI}. Assuming, for sake of 
simplicity, that the atom-field coupling is constant (we neglect thus the 
spatial dependence of the cavity mode amplitude), the duration of this Rabi 
pulse is $\pi/2\Omega$. With a proper choice for the atomic dipole phase and 
taking the energy of state $|g_s,1_a,0_b\rangle$ as the energy origin, the 
resulting atom-cavity state is, at time $t=\pi/2\Omega$:
\begin{equation}
|\Psi_1\rangle=\frac{1}{\sqrt 2}\left( |e_s,0_a\rangle+|g_s,1_a\rangle \right)|0_b\rangle\ .
\end{equation}
Atom $A_s$ and mode $M_a$ are then in a maximally entangled Einstein-Podolsky-Rosen 
(EPR) state \cite{EPR}. 

The atomic state, and thus its entanglement with $M_a$, are then copied onto mode 
$M_b$. Atom $A_s$ is tuned to resonance with $M_b$ ($\Delta=-\delta$) and undergoes 
a $\pi$ spontaneous emission pulse in this mode, performing the transformations 
$|g_s,0_b\rangle\rightarrow |g_s,0_b\rangle$ and
$|e_s,0_b\rangle\rightarrow i\exp(i\delta\pi/\Omega)|g_s,1_b\rangle$. The $i$ 
phase factor in the latter transformation is due to the fact that the polarization 
of mode $M_b$ is orthogonal to the one of mode $M_a$. The other phase factor 
originates in the $-\hbar\delta$ energy of state $|g_s,0_a,1_b\rangle$. This phase 
accumulates during the time $\pi/\Omega$ required for a $\pi$ Rabi rotation. Note 
that we neglect the light shifts produced by $M_a$ on the atom during its interaction with $M_b$. 

Atom $A_s$ finally ends up in $g$ and its state can be factorized out. The two 
modes end up at time $t=3\pi/2\Omega$ in the EPR pair state :
\begin{equation}
|\Psi_2(0)\rangle=\frac{1}{\sqrt 2}\left( e^{i\phi}|0_a,1_b\rangle+|1_a,0_b\rangle \right)\ ,
\label{EQ_ENTANGLED}
\end{equation}
where $\phi=\pi/2+\pi\delta/\Omega$. In terms of quantum information, this is a 
maximally entangled state of two qubits stored in the two cavity modes, the photon 
number states $|0\rangle$ and $|1\rangle$ being the qubits logical levels. At a 
later time $t$, the phase due to the energy difference $\hbar\delta$ between the 
two modes accumulates futher, for a time interval $t-3\pi/2\Omega$. The two-mode 
state becomes thus~:
$|\Psi_2(t)\rangle=\left( i\exp(-i\delta\pi/2\Omega)\exp(i\delta t)|0_a,1_b\rangle+|1_a,0_b\rangle \right)/\sqrt 2$. 

The modes entanglement is revealed by sending in $C$ the probe atom $A_p$ at 
time $t=T$. Atom $A_p$, initially in $g$, undergoes first a resonant $\pi$ Rabi 
pulse induced by the photon stored in $M_a$ during a time interval $\pi/\Omega$. 
This pulse copies the state of $M_a$ onto $A_p$ The resulting $A_p-M_a-M_b$ state 
is : $|\Psi_3\rangle=|0_a\rangle\left( i\exp(i\delta\pi/2\Omega) \exp(i\delta T)
|g_p,1_b\rangle-|e_p,0_b\rangle \right)/\sqrt 2$. Cavity mode $M_a$ ends up in 
$|0\rangle$ and factorizes out. 

Atom $A_p$ interacts then with $M_b$ for a $\pi/2$ Rabi pulse (duration $\pi/2\Omega$), 
performing the transformations $|g_p,1_b\rangle\rightarrow\exp(i\delta\pi/2\Omega)
(i|e_p,0_b\rangle+|g_p,1_b\rangle)/\sqrt{2}$ and $|e_p,0_b\rangle\rightarrow\exp
(i\delta\pi/2\Omega)(|e_p,0_b\rangle+i|g_p,1_b\rangle)/\sqrt{2}$. The final 
$A_p-M_b$ state is thus, within a global phase factor~:
\begin{eqnarray}
|\Psi_4\rangle=&&\frac{1}{2}\big(i|g_p,1_b\rangle(1-e^{i\delta T}e^{i\delta\pi/2\Omega})
\nonumber\\&&+|e_p,0_b\rangle (1+e^{i\delta T}e^{i\delta\pi/2\Omega})\big)\ .
\end{eqnarray}
The probability $P_e(T)$ for finding $A_p$ in $e$ is 
\begin{equation}
P_e(T)=\left(1+\cos(\delta T+\Phi)\right)/2\ ,\label{EQ_PE}
\end{equation}
with $\Phi=\pi\delta/2\Omega$. It oscillates between zero and one as a function of 
the time interval $T$. This oscillation reveals the coherent nature of the mode 
states superposition. 

The experimental set-up (figure \ref{FIG_SCHEME}(a)) has essentially been described 
in \cite{STEP,QRABI,EPRPAIR,MEMORY,QGATE,QND}. The rubidium circular levels $e$ and 
$g$ have principal quantum numbers 51 and 50 respectively ($e\rightarrow g$ transition 
frequency 51.1 GHz). The atoms are prepared in zone $B$ by a time-resolved excitation 
of a velocity-selected ($v=503\pm\,2$ m/s) atomic beam. Atoms $A_s$ and $A_p$ belong 
to two circular state samples separated by an adjustable time interval $T$. The 
position of these two samples at any time is known within $\pm 1$ mm, allowing us 
to control precisely and independently the Rabi pulses. Each atomic sample contains 
on the average 0.12 atoms. The probability for having two atoms in the same sample 
is thus low. Data acquisition selects events where a single atom is detected (in $D$) 
in each sample. The atomic path is enclosed in a 1.3~K cryostat protecting the atoms 
from the resonant blackbody field. The experimental sequence is repeated at a 600 s$^{-1}$ rate. 

The cavity $C$ is a Fabry Perot resonator made of two spherical niobium mirrors. The two 
orthogonally polarized TEM$_{900}$ modes, $M_a$ and $M_b$, have the same gaussian geometry 
(waist $w=6$ mm). The frequency splitting due a slight mirror shape anisotropy, 
$\delta/2\pi=128.3\pm 0.1$~kHz, is measured by auxiliary microwave transmission 
experiments. The photon damping times are $T_{r,a}=1$ ms and $T_{r,b}=0.9$ ms for 
$M_a$ and $M_b$ respectively. At thermal equilibrium, the modes contain a small thermal 
field originating from room-temperature microwave leaks (mean photon numbers 0.8 and 1 
for modes $M_a$ and $M_b$ respectively). Before the beginning of the experimental 
sequence, this thermal field is ``erased" down to less than 0.1 photon on average in 
each mode \cite{QND}. It then relaxes back to thermal equilibrium.

A d.c. voltage applied across the mirrors maintains the atomic orbital plane perpendicular 
to the cavity axis. The atoms are thus coupled in the same way to both modes, with a 
vacuum Rabi oscillation frequency $\Omega/2\pi=47$ kHz at cavity center.  In this static 
electric field, the atomic levels undergo a quadratic Stark effect (transition shift 
-255 kHz/(V/cm)$^2$). They are in resonance with modes $M_a$ and $M_b$ in electric fiels 
equal to 0.26 V/cm and 0.76 V/cm respectively. The atom-field interaction is interrupted 
by a 1.1 V/cm field, shifting the atomic line 150 kHz below the frequency of $M_b$ ($\Delta=-278$ kHz).

Atom $A_s$ enters the cavity at resonance with $M_a$. After a $\pi/2$ spontaneous emission 
pulse, $A_s$ is rapidly (within 1 $\mu$s) tuned to resonance with $M_b$ ($A_s$ is then 
3 mm before the cavity axis). During the next 12 $\mu$s (6 mm path), $A_s$ experiences 
a $\pi$ spontaneous emission pulse in $M_b$. The largest Stark field value is then 
applied, freezing the evolution. Atom $A_s$ is detected in the expected level $g$ in 
86\% of the cases. The departure from unity originates from detector errors and from 
the imperfections of the Rabi pulses. Sequences with $A_s$ detected in $e$ are 
discarded. Atom $A_p$ is resonant with $M_a$ when it enters $C$. The resonance condition 
is maintained until the atom fullfills the $\pi$ Rabi pulse condition. Atom $A_p$ then 
interacts with $M_b$ for a $\pi/2$ pulse and is finally detuned by a large field.

Figure \ref{FIG_FRINGES} presents the probability $P_e(T)$ for detecting $A_p$ in $e$ 
as a function of $T$. The data sets (a), (b), (c) and (d) correspond to four different 
windows in the $0\rightarrow 710 \ \mu$s time interval. The dots are experimental, 
with error bars reflecting the binomial detection statistics variance. The curves are 
sine fits. The frequency of these fits is the independently determined $\delta$ value. 
The four fits have independently adjustable contrasts and offsets. They share a common 
phase $\Phi$. Since the qualitative model of the experiment presented above does not 
take into account properly many important effects (spatial dependence of the cavity modes, 
light shifts...), $\Phi$ does not take the simple $\delta\pi/2\Omega$ value predicted 
by Eq. (\ref{EQ_PE}). We thus adjust $\Phi$ on the experimental data. 

We clearly observe the beat note between the two modes sharing a single photon at the 
expected frequency $\delta$. The contrast of the beat decreases with time, due to 
the cavity modes relaxation towards thermal equilibrium. At very long times, $P_e$ 
reaches a 30\% limit value, due to the absorption of the equilibrium thermal fields 
in both modes by atom $A_p$. Finally, we have checked that $P_e(T)$ is not modulated 
and reaches the same limiting 30\% value when atom $A_s$ is not sent.

This experiment shows that two cavity modes can be manipulated and coupled together 
according to pre-determined schemes. Many variants of this experiment could be 
implemented, opening new perspectives for quantum information manipulations in cavity 
QED experiments. For example, a quantum logic gate could be operated with the two modes 
as qubits. Mode $M_a$ and $M_b$ would play the role of the control and target qubits 
respectively. To implement the gate operation, an atom is sent across the cavity and 
undergoes a $\pi$ Rabi pulse in a one photon $M_b$ field. Its state becomes then a 
copy of the target. The atom is then detuned from the empty $M_b$ mode, while remaining 
non-resonant with $M_a$. The atomic coherence thus undergoes a phase shift which depends 
upon the photon number in $M_a$ \cite{QNDPROPOSAL}. In this way, the conditional dynamics 
of a quantum phase gate can be realized. The modified target state would finally be 
copied back onto $M_b$ by a resonant $\pi$ spontaneous emission pulse. The process can 
be generalized to couple more qubits. For example, the non-resonant interaction of an 
atom with both modes acting as joint control qubits leads naturally to three-qubit gates. 
Combining such gates implying two field modes and up to four atoms makes it possible to 
implement simple quantum algorithms \cite{FUMIKO}. Finally, the entangling scheme described 
in this paper could be applied to modes belonging to two separate cavities, realizing 
non-local field state entanglement \cite{MEYSTRE}. Such states could be used for teleportation 
of matter particle states \cite{TELEPORT}.

{\bf Acknowledgements} This work was supported by the Commission of the European Community 
and by the Japan Science and Technology Corporation (International Cooperative Research 
Project, Quantum Entanglement Project).

\begin{figure}
\caption{(a) Scheme of the experiment. (b) Temporal sequence for the two atoms. Qualitative 
plot of the detuning $\Delta$ between the atomic transition frequency and mode $M_a$ versus time.}
\label{FIG_SCHEME}
\end{figure}

\begin{figure}
\caption{Probability $P_e(T)$ for detecting $A_p$ in state $e$ as a function of the time 
interval $T$. Figures (a), (b), (c) and (d) correspond to four different time windows in 
the $0\rightarrow 710 \ \mu$s time interval. The dots are experimental, with errors bars 
reflecting the binomial statistics variance. The curves are sine fits.}
\label{FIG_FRINGES}
\end{figure}

\end{document}